\pdfoutput=1

\documentclass[11pt]{article}

\usepackage[final]{acl}

\usepackage{times}
\usepackage{latexsym}

\usepackage[T1]{fontenc}

\usepackage[utf8]{inputenc}

\usepackage{microtype}

\usepackage{inconsolata}

\usepackage{graphicx}

\usepackage{amsmath}
\usepackage{amssymb}
\usepackage{bm}
\usepackage{multirow}
\usepackage{booktabs}
\usepackage{algorithm}
\usepackage{algorithmic}

\title{STARS: A Unified Framework for Singing Transcription, Alignment, and Refined Style Annotation}

\makeatletter
\def\@fnsymbol#1{%
  \ensuremath{%
    \ifcase#1\or \dagger\or *\or \ddagger\or
    \mathsection\or \mathparagraph\or \|\or **\or \dagger\dagger
    \or \ddagger\ddagger \else\@ctrerr\fi%
  }%
}
\makeatother

\author{%
Wenxiang Guo\thanks{Equal contribution}\quad
Yu Zhang\footnotemark[1]\quad
Changhao Pan\footnotemark[1]\quad
Zhiyuan Zhu\quad \\
\textbf{Ruiqi Li} \quad 
\textbf{Zhetao Chen} \quad
\textbf{Wenhao Xu}\quad
\textbf{Fei Wu} \thanks{Corresponding Author}\quad
\textbf{Zhou Zhao} \footnotemark[2]\\
 Zhejiang University \\
  \texttt{\{guowx314,yuzhang34,zhaozhou\}@zju.edu.cn} \\
}

\begin{document}
\maketitle
\begin{abstract}
Recent breakthroughs in singing voice synthesis (SVS) have heightened the demand for high-quality annotated datasets, yet manual annotation remains prohibitively labor-intensive and resource-intensive. Existing automatic singing annotation (ASA) methods, however, primarily tackle isolated aspects of the annotation pipeline. To address this fundamental challenge, we present STARS, which is, to our knowledge, the first unified framework that simultaneously addresses singing transcription, alignment, and refined style annotation. Our framework delivers comprehensive multi-level annotations encompassing: (1) precise phoneme-audio alignment, (2) robust note transcription and temporal localization, (3) expressive vocal technique identification, and (4) global stylistic characterization including emotion and pace. The proposed architecture employs hierarchical acoustic feature processing across frame, word, phoneme, note, and sentence levels. The novel non-autoregressive local acoustic encoders enable structured hierarchical representation learning. Experimental validation confirms the framework's superior performance across multiple evaluation dimensions compared to existing annotation approaches. Furthermore, applications in SVS training demonstrate that models utilizing STARS-annotated data achieve significantly enhanced perceptual naturalness and precise style control. This work not only overcomes critical scalability challenges in the creation of singing datasets but also pioneers new methodologies for controllable singing voice synthesis. Audio samples are available at \url{https://gwx314.github.io/stars-demo/}.

\end{abstract}

\section{Introduction}
Automatic Singing Annotation (ASA) constitutes the computational process of extracting key vocal features from singing recordings, encompassing phonetic transcriptions (phoneme alignment), MIDI note parameters (pitch/duration), and stylistic attributes (emotion/technique). As the cornerstone of modern singing voice synthesis (SVS) systems, ASA provides fine-grained, multi-level annotated data for training expressive and controllable singing synthesis models. 
While recent breakthroughs in generative models and controllable SVS frameworks \citep{guo2025techsinger,zhang2024stylesinger,zhang2024tcsinger} have dramatically improved the quality of generated singing voices, they have also paradoxically exposed a critical bottleneck: the scarcity of high-quality annotated singing corpora.

Traditional annotation workflows require labor-intensive manual processing by audio engineers and musicians, making large-scale dataset creation both costly and time-consuming. While some open-source singing datasets such as OpenCpop \citep{wang2022opencpop} and VocalSet \citep{wilkins2018vocalset} have attempted to alleviate this burden, their annotations are limited to basic phonetic or vocal technique information. Recent datasets like GTSinger \citep{zhang2024gtsinger} have made significant progress by incorporating a wider range of annotations—from basic phoneme and note annotations to various singing techniques and global styles. However, the volume of data remains insufficient compared to the scale of speech corpora.

\begin{figure*}[t]
\centering
\includegraphics[width=0.9\textwidth, trim={10mm 70mm 10mm 40mm}, clip]{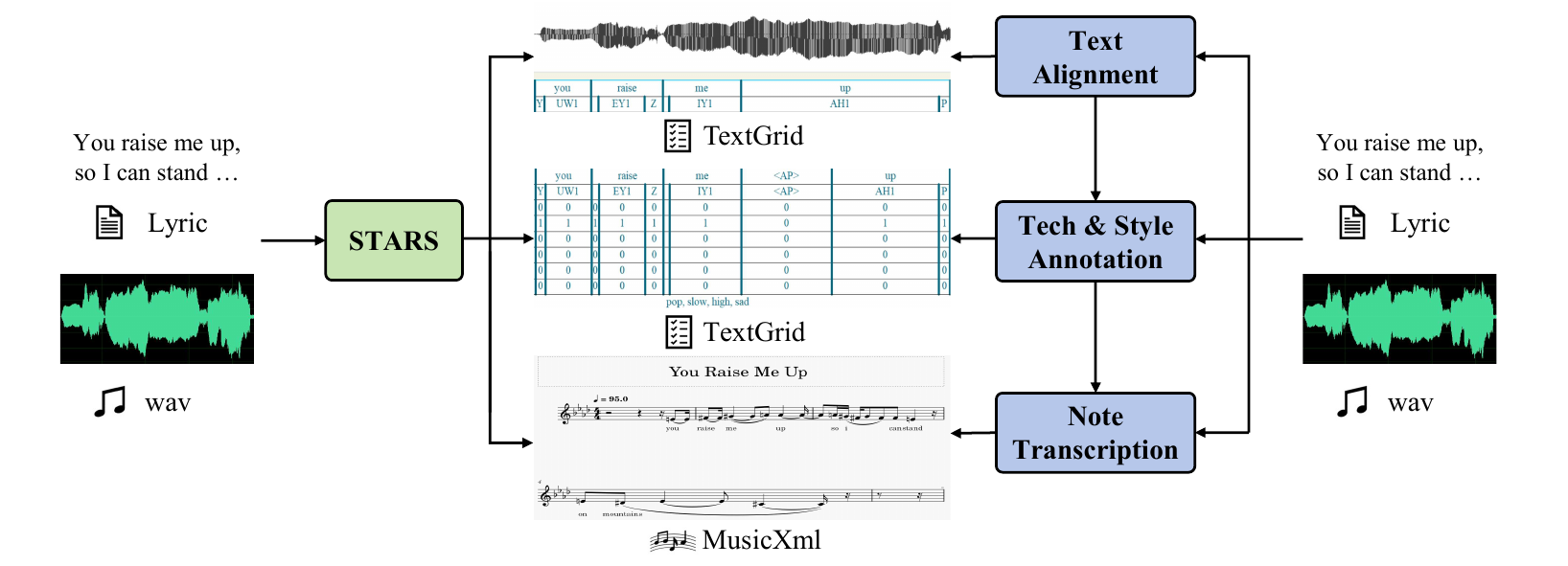}
\caption{STARS vs. Traditional Stepwise Pipeline. 
Conventional stepwise processing requires sequential execution of text alignment, note transcription, and manual technique and style annotation, with error propagation across cascaded modules. 
STARS establishes unified acoustic-linguistic modeling for simultaneous phoneme, MIDI, technique, and style prediction, eliminating error accumulation via end-to-end joint optimization.
}
\label{fig: head}
\end{figure*}

Modern SVS systems require multi-level annotation precision across four key dimensions: (1) microsecond-aligned phoneme boundaries for prosody modeling; (2) accurate MIDI note timing/pitch for melody preservation; (3) phone-level vocal technique recognition (e.g., vibrato, falsetto); and (4) global stylistic attributes (emotion, pace). As shown in Figure~\ref{fig: head}, traditional solutions employ fragmented toolchains—combining ASR systems like WhisperX \citep{bain2023whisperx} and Qwen-Audio \citep{chu2024qwen2} for lyric transcription, MFA \citep{mcauliffe2017montreal} for forced alignment, and pitch trackers like VOCANO \citep{hsu2021vocano} and MusicYOLO \citep{wang2022musicyolo}. This patchwork approach introduces cascading errors from tool mismatches while failing to capture expressive vocal styles. Such disjointed processes hinder the creation of large, high-quality annotated singing datasets necessary for cutting-edge SVS models.

To overcome these challenges, we propose STARS, a unified framework for multi-level singing voice annotation that streamlines the entire process. STARS offers three key innovations: 
(1) A multi-level architecture for extracting singing information at various granularities, covering frame, word, phoneme, note, and sentence levels; 
(2) A local acoustic encoder that works together with a CMU encoder employing a U-Net architecture with Conformer blocks and FreqMOE, and with a vector quantization, to extract acoustic features at multiple hierarchical levels; and 
(3) A multi-task automated annotation pipeline that sequentially predicts phoneme boundaries, note boundaries, pitch, phone-level techniques, and global stylistic attributes. We demonstrate the effectiveness of STARS through comprehensive evaluations across multiple ASA tasks. Our framework achieves superior performance in phoneme alignment accuracy and note prediction precision. When applied to SVS model training, STARS-annotated data yields significant improvements in synthesized voice naturalness and style control accuracy.

\begin{itemize}
     \item We propose STARS, the first unified framework for Singing Transcription, Alignment, and Refined Style Annotation.
 
    \item We design a five-level unified architecture with specialized acoustic encoders for hierarchical feature filtering and extraction.

    \item We implement parallel prediction strategies for phoneme/note boundaries and pitch estimation, enhanced through phone-level technique and global style detection.
    
    \item Through training SVS models with our annotations, we demonstrate the practical utility of STARS in achieving superior singing vocal naturalness and precise style control.

 \end{itemize}
 
\section{Related Works}

\subsection{Singing Voice Synthesis}
Singing Voice Synthesis (SVS) aims to generate expressive singing voices from musical scores. Early models such as XiaoiceSing \citep{lu2020xiaoicesing} adopt non-autoregressive acoustic architectures inspired by FastSpeech \citep{ren2019fastspeech}. Subsequent work, ViSinger \citep{zhang2022visinger}, employs VITS \citep{kim2021conditional} to establish an end-to-end SVS framework. Generative adversarial networks  \citep{wu2020adversarially, huang2022singgan} and diffusion models \citep{liu2022diffsinger} have also been applied to enhance audio fidelity. Controllable SVS focuses on manipulating vocal attributes including timbre, emotion, and style to achieve more expressive performances. \citet{resna2023multi} develops a multi-singer framework for cross-voice synthesis. Muse-SVS \citep{kim2023muse} enables precise control over pitch, energy, and phoneme duration to express different emotional intensities. Prompt-Singer \citep{wang2024prompt} introduces natural language prompts to achieve fine-grained control over singing voices. To mitigate data scarcity, DeepSinger \citep{ren2020deepsinger} constructs a large-scale corpus by mining singing data from online sources. Additionally, OpenCpop \citep{wang2022opencpop} and GTSinger \citep{zhang2024gtsinger} provide publicly available corpora with manually annotated singing recordings. Nevertheless, the limited availability of high-quality singing data remains a critical bottleneck compared to speech resources.

\subsection{Automatic Singing Annotation}

Automatic Singing Annotation (ASA) includes tasks such as lyric alignment, note estimation and segmentation, and vocal technique and style annotation. 
The MFA \citep{mcauliffe2017montreal} is a conventional approach for lyric alignment.
However, singing voice alignment remains challenging due to the large variations in phoneme durations and rhythmic structures. 
Several studies \citep{wang2023adapting, huang2022improving} adopt Viterbi forced alignment \citep{forney1973viterbi} to improve the accuracy of aligning posteriograms with lyrics.
For note estimation, VOCANO \citep{hsu2021vocano} and MusicYOLO \citep{wang2022musicyolo} directly predict the pitch and duration of musical notes. 
ROSVOT \citep{li2024robust} incorporates phoneme boundary priors to achieve MIDI note prediction. 
SongTrans \citep{wu2024songtrans} builds upon the Whisper model and adopts a hybrid autoregressive and non-autoregressive framework for phoneme and note annotation. 
MusCaps \citep{manco2021muscaps} leverages a language model to generate music captions, but its effectiveness is limited for singing voice recordings without background music.
Despite these advancements, existing ASA methods remain fragmented, requiring separate models and manual integration.

\section{STARS}
\subsection{Problem Formulation}
Typically, a Mel-spectrogram $\mathbf{M} \in \mathbb{R}^{T \times F}$ is derived from the audio signal using the short-time Fourier transform (STFT), where $T$ denotes the number of frames and $F$ the number of frequency bins. Let the phoneme sequence be represented as $p = [p_1, p_2, \dots, p_{L_p}]$, where $L_p$ is the number of phonemes. In addition to predicting the phoneme sequence, the model is required to predict the corresponding boundaries for each phoneme, represented as $ph\_bd = [pbd_1, pbd_2, \dots, pbd_T]$, where $pbd_i = 1$ indicates that frame $i$ is a phoneme boundary and $0$ otherwise. Furthermore, for each phoneme, the model predicts a set of singing techniques. For each technique $i \in \{1,\dots,9\}$, a binary sequence $tech^i = [t^i_1, t^i_2, \dots, t^i_{L_p}]$ is predicted, where $t^i_j = 1$ indicates the presence of technique $i$ at the $j$-th phoneme and $t^i_j = 0$ indicates its absence. Based on the predicted phoneme and word boundaries, note boundaries are predicted as $note\_bd = [nbd_1, nbd_2, \dots, nbd_T]$. The model also predicts the note pitch sequence $c = [c_1, c_2, \dots, c_{L_n}]$, where $L_n$ denotes the number of notes. In addition to the phoneme, note, and technique predictions, the model is required to predict global sentence-level attributes $\mathbf{g}$.

\begin{figure*}[t]
\centering
\includegraphics[width=0.9\textwidth, trim={0mm 20mm 0mm 5mm}, clip]{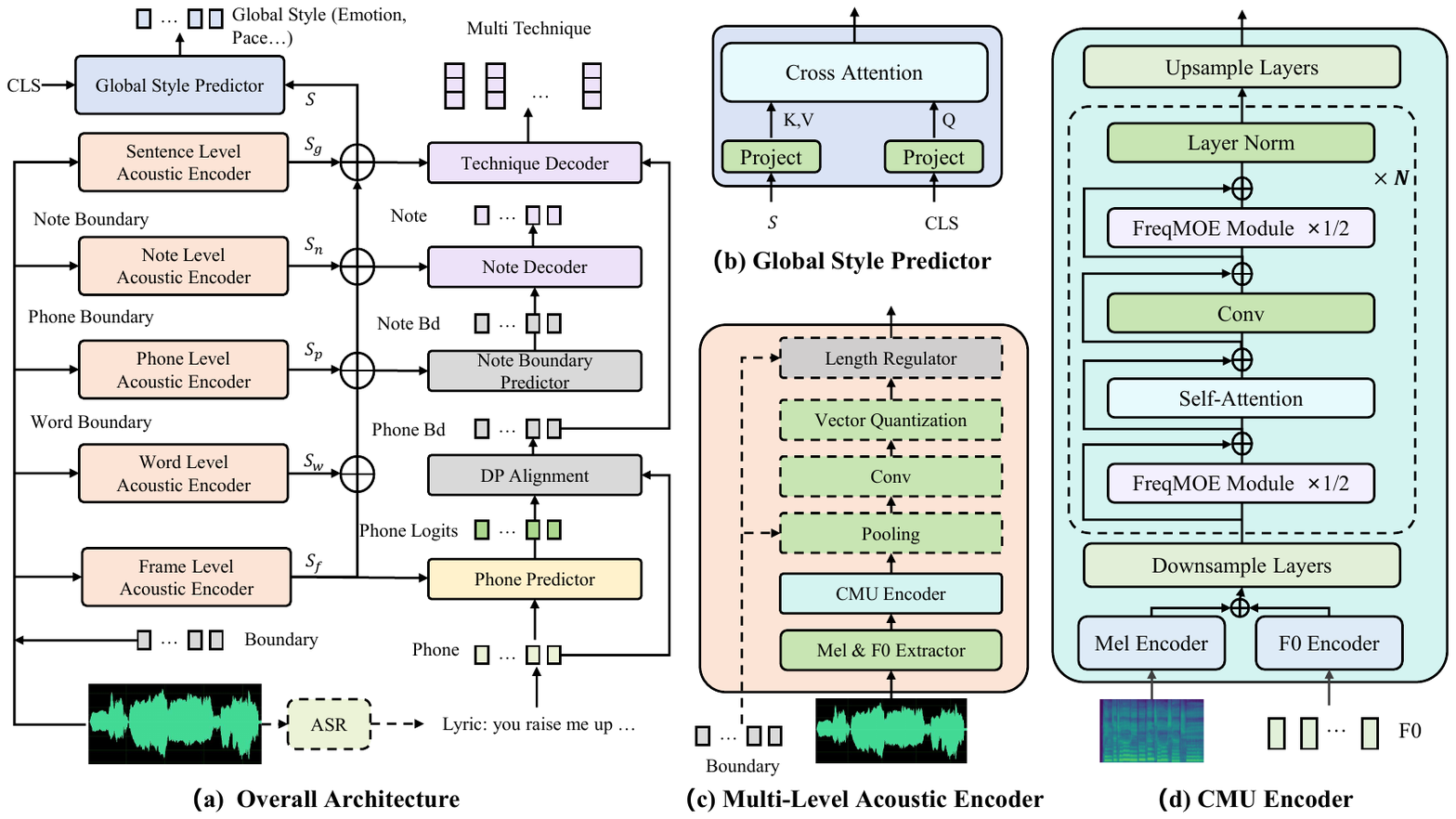}
\caption{
The overall architecture of STARS. 
(a) The unified multi-level framework, which integrates singing lyric alignment, note transcription, technique prediction, and global style prediction. 
(b) The Global Style Predictor, where [CLS] tokens are used as queries, and $\mathcal{S}$ represents the keys and values. 
(c) The multi-level acoustic encoder extracts features at each level, with optional pooling based on boundary segmentation.
(d) The CMU Encoder, employing a U-Net architecture with Conformer blocks and FreqMOE for efficient audio feature extraction.
}
\label{fig: arch}
\end{figure*}
\subsection{Overview}
Figure~\ref{fig: arch} illustrates the overall architecture of STARS. The input to our system consists of Mel-spectrograms and F0 extracted from the audio signal, and the corresponding lyrics can be obtained using ASR models such as Whisper \citep{radford2023robust}. To improve the robustness of our model, we follow the approach of ROSVOT \citep{li2024robust} by adding realistic noise from the MUSAN dataset~\cite{snyder2015musan} to the input audio and injecting Gaussian noise into the extracted F0 contour.
In this section, we first describe the unified multi-level framework. To capture multi-level acoustic and stylistic information, we design a hierarchical architecture that spans five levels: Frame, Word, Phone, Note, and Sentence. Each level shares the same backbone while employing slightly different methods to efficiently extract features at varying granularities.
Next, we explain how all sub-tasks are completed in a single forward pass. To obtain the Phone and Word boundaries, we first predict the frame-level phoneme logits from the features extracted by the Frame-level encoder. We then apply Viterbi forced alignment to determine the phoneme and word boundaries. For note boundaries, we utilize features from the previous three levels to predict the note boundaries. Having extracted features at the Note level, we can predict the corresponding note pitch.
Finally, leveraging the information from all five levels, we predict various singing techniques and global attributes.

\subsection{Unified Multi-Level Framework}
\label{sec:level}
To achieve unified annotation predictions at multiple levels within a single model, we design a Unified Multi-Level Framework consisting of five hierarchical levels: Frame, Word, Phone, Note, and Sentence. Each level extracts acoustic features at different granularities. The framework employs a shared acoustic encoder across all levels to enable efficient feature extraction. We first design a highly efficient CMU Encoder to extract features, followed by multi-granularity pooling operations. Vector quantization serves as a bottleneck~\cite{van2017neural} to eliminate irrelevant information. We then use the boundary to expand the extracted features to the frame length $T$.

The CMU Encoder module utilizes a U-Net-based architecture for Mel-spectrogram downsampling while preserving spectral details through skip connections during upsampling. To capture both long-term and short-term dependencies in the time dimension, we integrate the Conformer architecture~\cite{gulati2020conformer}, which demonstrates superior performance in the ASR tasks. For enhanced frequency analysis, we design the FreqMOE module that partitions the frequency dimension into $K$ equal frequency bands and applies the specialized experts to distinct frequency bands. We then concatenate the output embedding chunks. Specific details are provided in the Appendix~\ref{sec: moe}.

For the pooling module, no pooling is applied to the frame-level features, as this level represents the finest granularity. For each intermediate level, we utilize the corresponding boundary information to perform segmentation, followed by average pooling within these boundaries to obtain dynamic feature sequences. At the sentence level, we apply global average pooling to the CMU Encoder outputs to obtain the holistic representation.

For hierarchical representation learning, we apply vector quantization to the intermediate features from all levels except at the Frame and Sentence levels. Let $\mathbf{S}_l \in \mathbb{R}^{L \times D}$ denote the input latent embeddings for level $l$, where $L$ is the sequence length and $D$ is the feature dimension. Each level maintains a codebook $\mathbf{q}_l \in \mathbb{R}^{K \times D}$ containing $K$ latent embeddings. Following~\cite{van2017neural}, we also apply a commitment loss to ensure that the representation sequence commits to an embedding and to prevent the output from growing:

\begin{equation}
    \mathcal{L}_{\text{commit}} = \|\mathbf{z}_l(\mathbf{S}_l) - \text{sg}[\mathbf{q}_l]\|_2^2,
\end{equation}
where \(z_l(.)\) is the vector quantization module for level \(l\), and \(\operatorname{sg}\) denotes the stop gradient operator.

In the Length Regulator module, at each level \(l \in \{\text{word}, \text{phone}, \text{note}\}\), we use the boundary information to determine the frame length of each segment. To align hierarchical representations, we repeat each embedding according to the length of the segment. At the Sentence level, we extend the single embedding to match the frame length $T$.

\subsection{Lyric Alignment}
\label{sec: alignment}
Phonemes are the smallest phonetic units and the most commonly used tokens in singing voice synthesis. To align phonemes with the audio, we input the frame-level extracted features, denoted as \(\mathcal{S}_f\), into the phone predictor. This predictor generates predictions for the phoneme being sung at each frame, along with indications of whether a frame corresponds to a phoneme boundary. During training, we optimize the phone predictions using cross-entropy loss, and we also apply Connectionist Temporal Classification (CTC) loss~\cite{graves2006connectionist} to further improve phoneme alignment.

\begin{equation}
    L_{\text{CE}} = -\sum_{k=1}^{P} y_k \log(p_k),
\label{eq: ce}
\end{equation}
\begin{equation}
    \mathcal{L}_{\text{CTC}} = -\log \sum_{\pi \in \mathcal{B}^{-1}(\mathbf{y})} P(\pi|\mathbf{X}),
\end{equation}
where \(P\) is the total number of phonemes, and \(\pi = [\pi_1, \dots, \pi_T]\) represents an alignment path with \(\pi_t \in \mathcal{V} \cup \{\text{blank}\}\) (where \(\mathcal{V}\) is the total phoneme vocabulary). Additionally, we use BCE loss to predict the phone boundaries.

During inference, we employ the Viterbi forced alignment~\cite{forney1973viterbi} method to align the lyric phoneme sequence with the phone probability distribution at each frame. This process provides the phoneme boundaries, and through the phone-to-word relationship, we further determine the word boundaries. Specific details of the inference algorithms are provided in the Appendix~\ref{sec: appalign}.

\subsection{Note Transcription and Alignment}
\label{sec: midi}
To obtain note boundaries, we first fuse features from the frame, word, and phone levels and feed them into a note boundary predictor. The training loss for note boundary prediction is defined as the BCE loss, similar to phone boundary prediction. During inference, since the word boundaries overlap with the note boundaries, we employ these boundaries as constraints when generating the final note boundaries.
For pitch prediction at the note level, we integrate the aggregated features from the frame, word, and phone levels with note-level features. Specifically, for each note \(j\), we denote its feature sequence as \(\mathbf{S}_n^{j} \in \mathbb{R}^{L_j \times D}\), where \(L_j\) is the number of frames in the note segment and \(D\) is the feature dimension. We then use the note decoder to predict the pitch of each note. Inspired by CIF~\cite{dong2020cif}, we also compute the weight vector for each note as \(W_n = \mathbf{S}_n^{j} \mathbf{W}_a\), where \(\mathbf{W}_a\) is a learnable projection matrix and \(W_n \in \mathbb{R}^{L_j \times 1}\). We then obtain the aggregated note representation \(c_j\) via a weighted average:

\begin{equation}
    c_j = \sum_{t=1}^{L_j} \mathbf{w}_n(t) \, \mathbf{S}_n^{j}(t),
\end{equation}
where \(\mathbf{w}_n(t)\) denotes the weight at frame \(t\). 

We then use the pitch predictor to compute the logits for the \(P\) pitch categories \(\hat{p}_j = c_j\,\mathbf{W}_O\), where \(\mathbf{W}_O \in \mathbb{R}^{D \times P}\). The pitch predictor is also optimized using cross-entropy loss.

\subsection{Technique and Global Style Predictor}
\label{sec: style}
To predict the possible techniques for each phoneme, we treat this task as a multi-task, multi-label binary classification problem. The technique prediction head outputs predictions across nine categories: mixed, falsetto, strong, weak, glissando, breathy, bubble, vibrato, and pharyngeal. For each phoneme \(j\), we predict the \(i\)-th technique \(tech_j^i\), where \(1\) indicates the presence of the technique and \(0\) indicates its absence. Features from the Frame, Word, Phone, Note, and Sentence levels are input into the model, with the previously obtained phone boundaries used as references. We apply the same attention-based weighted average strategy used for pitch prediction to predict each technique sequence. Binary cross-entropy (BCE) loss is used to optimize each technique prediction module.

For the global attributes of the sentence, including language, gender, emotion, pace, and range, we treat these as multi-class classification tasks. To predict these global attributes, we introduce five [CLS] tokens, each corresponding to one of the tasks. These tokens are used as queries \(\mathcal{H}_{c}\), while the sum of the frame-level features from all levels, i.e., \(\mathcal{S}_f\), \(\mathcal{S}_w\), \(\mathcal{S}_p\), \(\mathcal{S}_n\), and \(\mathcal{S}_g\), serves as the key and value in the cross-attention mechanism~\cite{vaswani2017attention}. Positional encoding embeddings are added to the features to determine the position of each token. The formulation is as follows:

\begin{equation}
\begin{aligned}
    &\mathcal{S} = \mathcal{S}_f + \mathcal{S}_w + \mathcal{S}_p+\mathcal{S}_n+\mathcal{S}_g, \\
    &\text{Attention} (\mathcal{H}_{c}, \mathcal{S}, \mathcal{S})
     = \text{Softmax} \left( \frac{\mathcal{H}_{c} \mathcal{S}^T}{\sqrt{D}} \right) \mathcal{S},
\end{aligned}
\end{equation}
where \(D\) is the dimension of the query, key and value.
Then, we predict each task’s category \(g_i \in \mathbb{R}^{C_i}\), where \(C_i\) is the number of categories for the \(i\)-th attribute. Cross-entropy (CE) loss is used to train and optimize the global style predictors.

\subsection{Training and Inference Procedures}
During training, we leverage ground truth phone, word, and note boundaries to guide the model. The final loss function consists of the following components: 
1) \(L_{\text{ph}}\): A combination of phone-level CTC loss and CE loss; 
2) \(L_{\text{pbd}}\) and \(L_{\text{nbd}}\): Boundary prediction losses for the phone and note, respectively, optimized using BCE loss; 
3) \(L_{\text{pi}}\): Pitch prediction loss, calculated as the CE loss between the ground truth pitch and the predicted pitch; 
4) \(L_{\text{tech}}\): Technique prediction loss, calculated as the CE loss; 
5) \(L_{g}\): The global attribute prediction loss, optimized using CE loss; and 
6) \(L_c\): The commitment loss, which constrains the vector quantization layer.

During inference, we first obtain the phonemes and words from the input lyrics (or ASR model-generated lyrics) and use the frame-level predicted phoneme logits with Viterbi forced alignment to determine the phoneme boundaries. Note boundaries are then predicted by feeding fused features from the frame, word, and phone levels into the note boundary predictor. The note-level features, together with the fused features, are provided as input to the note-level acoustic encoder and the note decoder to predict the note pitch. Next, features from all levels, along with the phoneme boundaries, are used to predict the techniques for each phoneme. Finally, the global attributes are predicted using the aggregated features from all levels.

\section{Experiments}

\subsection{Experimental Setup}
\begin{table*}[t]
\centering
\begin{minipage}{0.48\textwidth}
\centering
\begin{tabular}{l|cc}
\toprule
\rule{0pt}{10pt} \bfseries{Method} & \bfseries{BER} $\downarrow$ & \bfseries{IOU} $\uparrow$ \\
\hline
MFA & 40.3 & 56.8 \\
SOFA & 20.9 & 80.0 \\
\midrule
\bf STARS (ours) & \bf 18.6 & \bf 80.9 \\
\bottomrule
\end{tabular}
\caption{Results for lyric alignment.}
\label{tab: lyric_alignment}
\end{minipage}
\hfill
\begin{minipage}{0.48\textwidth}
\centering
\begin{tabular}{l|cc}
\toprule
\rule{0pt}{10pt} \bfseries{Method} & \bfseries{COnPOff(F)} $\uparrow$ & \bfseries{RPA} $\uparrow$ \\
\hline
VOCANO & 50.2 & 76.6 \\
ROSVOT & 70.2 & 83.8 \\
\midrule
\bf STARS (ours) & \bf 71.0 & \bf 86.7 \\
\bottomrule
\end{tabular}
\caption{Results for note transcription and alignment.}
\label{tab: note_transcription}
\end{minipage}
\end{table*}
\begin{table*}[t]
\centering
\small
\begin{tabular}{l|c|ccccccccc|cc}
\toprule
\multirow{2}{*}{\bfseries{Setting}} & \multirow{2}{*}{\bfseries{Metric}} & \multicolumn{11}{c}{\textbf{Phone-level Technique and Global Style Prediction}} \\
& & {BUB} & {BRE} & {PHA} & {VIB} & {GLI} & {MIX} & {FAL} & {WEA} & {STR} & \bfseries{TEC} & \bfseries{STY}\\
\midrule
\multirow{2}{*}{GTSinger} 
& F1   & 46.9 & 68.7 & 88.7 & 95.7 & 78.5 & 61.5 & 33.2 & 37.2 & 82.5 & 67.3 & - \\
& ACC  & 31.5 & 73.2 & 75.7 & 99.3 & 78.9 & 93.9 & 40.8 & 17.4 & 95.3 & 65.9 & - \\
\midrule
\multirow{2}{*}{\textbf{STARS}} 
& F1    & 71.7 & 66.9 & 85.0 & 65.5 & 72.3 & 74.7 & 93.5 & 90.3 & 99.4 & \bf 79.9 & - \\
& ACC   & 97.8 & 88.8 & 95.4 & 96.7 & 84.1 & 81.9 & 94.7 & 90.4 & 93.9 & \bf 91.5 & 68.0 \\
\bottomrule
\end{tabular}
\caption{
The objective results of phone-level technique prediction. 
The singing techniques include BUB (bubble), BRE (breathy), PHA (pharyngeal), VIB (vibrato), GLI (glissando), MIX (mixed), FAL (falsetto), WEA (weak), and STR (strong). The "TEC" column represents the average metrics calculated across all the singing techniques, while the "STY" column represents the average metrics calculated across all the global style attributes.
}
\label{tab: tech}
\end{table*}

\begin{table*}[t]
\centering
\begin{tabular}{l|cc|cc|cc|c}
\toprule
\rule{0pt}{10pt} \bfseries{Method}& \bfseries{BER} $\downarrow$ & \bfseries{IOU} $\uparrow$ & \bfseries{COnPOff(F)} $\uparrow$ & \bfseries{RPA} $\uparrow$  & \bfseries{T-F1} $\uparrow$ & \bfseries{T-ACC} $\uparrow$  & \bfseries{S-ACC} $\uparrow$ \\
\hline
\bf STARS & 18.6 & \bf 80.9 & \bf 71.0 & \bf 86.7 & \bf 79.9 & \bf 91.5 & 68.0 \\
\midrule
w/o CTC & 19.1 & 80.0 & 70.9 & 86.7 & 79.7 & 89.8 & 66.8 \\
w/o VQ & \bf 18.3 & 80.9 & 70.7 & 86.4 & 76.3 & 90.4 & 65.3  \\
w/o MOE & 18.9 & 80.5 & 70.1 & 86.5 & 77.4 & 90.0 & 69.8 \\
Conv & 20.1 & 78.3 & 66.4 & 82.7 & 62.9 & 89.6 & 66.6 \\
Bilingual & 19.4 & 75.6 & 68.1 & 86.2 & 76.8 & 91.4 & \bf 70.4 \\
\bottomrule
\end{tabular}
\caption{
The ablation results for different sub-tasks. T-F1 is the average F1-score for all singing techniques, T-ACC is the average accuracy for all singing techniques, and S-ACC is the average accuracy for all global style attributes.
}
\label{tab: abl}
\end{table*}
\subsubsection{Dataset and Process}
\label{sec: data}
The dataset used in our experiments includes the Chinese and English subsets of GTSinger~\citep{zhang2024gtsinger}. This dataset provides alignments and annotations in TextGrid files, which include word boundaries, phoneme boundaries, phoneme-level annotations for six techniques (mixed, falsetto, pharyngeal, glissando, vibrato, breathy), and global style labels such as emotion, pace, and pitch range. Additionally, we have collected and annotated a 30-hour Chinese dataset featuring two singers and four technique annotations (mixed, falsetto, strong, weak, breathy, bubble) at both the phoneme and sentence levels. To train our automated annotation model, we reserve 30 songs containing various techniques and global styles as the validation and test sets. To ensure the robustness of the model, we augment the dataset by adding noise from the MUSAN noise corpus~\cite{snyder2015musan}. For Chinese lyrics, we use the pypinyin tool\footnote{\url{https://github.com/mozillazg/python-pinyin}} to phonemize the text, while for English lyrics, we follow the ARPA\footnote{\url{https://en.wikipedia.org/wiki/ARPABET}} standard for phoneme transcription.

\subsubsection{Implementation Details}
\label{sec: imple}
The singing audio recordings are sampled at 24 kHz, with a window size of 512 samples, a hop size of 128, and 80 mel bins for Mel-spectrogram extraction. We use a pre-trained RMVPE~\cite{wei2023rmvpe} model to extract the F0 contours. The U-Net backbone consists of four downsampling and upsampling layers, with a total downsampling factor of 16×. The Conformer module includes two layers, and the FreqMOE consists of four experts. In this experiment, the model is trained for 150k steps using an NVIDIA 4090 GPU. Further implementation details are provided in Appendix~\ref{sec: app2model}.

\subsubsection{Evaluation Details}
\label{sec: eval}
For lyric alignment, we evaluate performance using two metrics: Boundary Error Rate (BER) and Intersection Over Union (IOU) score. BER measures the proportion of misplaced boundaries within 20ms tolerance. The IOU score is defined as the ratio of the duration of the overlapping segment between two notes to the duration of the combined time span covered by both notes. For note transcription, we use the \texttt{mir\_eval} library~\cite{raffel2014mir_eval} and apply the metrics COnPOff (correct onset, pitch, and offset) proposed in~\cite{molina2014evaluation} and Raw Pitch Accuracy (RPA) for overall note pitch prediction performance. For phone-level technique and global style recognition, we use objective metrics including F1 score and accuracy to evaluate the phone-level technique predictor, and accuracy for the global style detector. The results are multiplied by 100 for better readability. Further details are provided in Appendix~\ref{sec: appendix4metr}.

\subsubsection{Baseline Models}
To evaluate our approach, we compare it with several baseline systems across different sub-tasks. We conduct the comparison using only Chinese data, and additionally, we test the model's performance on multilingual data by combining both Chinese and English datasets. For the phoneme and singing audio alignment, we consider:
1) Montreal Forced Align (MFA): a tool that aligns orthographic and phonological forms by leveraging a pronunciation dictionary to time-align transcribed audio files;
2) SOFA~\footnote{\url{https://github.com/qiuqiao/SOFA}}: a forced alignment tool designed specifically for the singing voice.
For note alignment and transcription, we compare our model’s performance with the results reported in ROSVOT, including two strong baselines:1) VOCANO: a note transcription framework developed for the singing voice in polyphonic music;
2) ROSVOT: a model that employs a multi-scale architecture for automatic singing transcription. We compare with the variant without word boundary condition.
For phone technique prediction, we use GTSinger's technique predictor as the baseline.

\subsection{ASA Results}
\subsubsection{Lyric Alignment and Note Transcription}
In Table~\ref{tab: lyric_alignment}, we observe that for the lyric alignment task, when comparing our model with the other two models, STARS achieves the best performance in both the BER and IOU metrics, indicating its ability to accurately predict phoneme boundary information.
In Table~\ref{tab: note_transcription}, we see that for the note alignment and transcription task, our model outperforms the baseline models in both COnPOff(F) and RPA metrics, demonstrating its sensitivity to note boundaries and pitch.
Notably, while the other models are designed to handle only a single task, our model efficiently handles both lyric alignment and note transcription tasks simultaneously. This demonstrates the versatility and effectiveness of STARS in performing multiple singing annotation tasks, making it highly suitable for foundational automatic singing annotation applications.

\subsubsection{Technique and Global Style Prediction}
As Table~\ref{tab: tech} shows, for the recognition of the nine vocal techniques, our experimental results outperform GTSinger. No individual technique shows a significantly low recognition accuracy. The average F1-score and accuracy for all techniques far exceed the GTSinger benchmark, demonstrating our model's ability to accurately detect and annotate multiple techniques at the phoneme level. Also, as the last column of the table indicates, our model achieves high accuracy in recognizing global style attributes, further showcasing its effectiveness in capturing the overall stylistic features of singing audio. For detailed scores of each style attribute, see the Appendix~\ref{sec: appendix2biling}.
In summary, our model effectively detects expressive information, and the generated labels can be used in various controllable expressive singing voice synthesis tasks.

\subsubsection{Ablation Study}
In this section, we conduct ablation experiments to evaluate the contributions of different components in our model. We test the following variants:
1) w/o CTC: the model without the CTC loss;
2) w/o VQ: the model without vector quantization;
3) w/o MOE: the model without the FreqMOE strategy;
4) Conv: the model with the Conformer module replaced by a convolutional architecture;
5) Bilingual: the model trained on bilingual datasets.

As shown in Table~\ref{tab: abl}, several conclusions can be drawn. Comparing the first row with the w/o CTC variant, we observe improvements of 0.5 and 0.9 in the BER and IOU metrics for lyric alignment, respectively, along with enhanced performance in the technique recognition task, which is sensitive to phoneme boundaries, while the note recognition metrics remain similar. These results indicate that CTC loss notably enhances phoneme alignment.
When VQ is omitted for the phone, note, and word levels, the note, technique, and style recognition tasks show improved performance. For phoneme alignment, the lighter model with reduced VQ loss training further improves phoneme boundary detection.
Comparisons of the w/o MOE and Conv variants reveal a drop in performance across all metrics, demonstrating the effectiveness of Finally, experiments on bilingual datasets confirm that our model operates effectively in a multilingual setting. Detailed results for single-language and bilingual experiments can be found in Appendix~\ref{sec: appendix2biling}.

\subsection{Singing Voice Synthesis}
\label{sec: svs}
\begin{table}[t]
\centering
\small
\begin{tabular}{ll|cc}
\toprule
\bfseries{Train} & \bfseries{Infer} & \bfseries{MOS-Q} $\uparrow$ & \bfseries{MOS-C} $\uparrow$\\
\midrule
\bfseries{GT}      & \bfseries{GT}     & 3.98 $\pm$ 0.09   & 4.05 $\pm$ 0.09 \\
\bfseries{GT}      & \bfseries{Pred}     & 3.91 $\pm$ 0.07   & 3.95 $\pm$ 0.08 \\
\bfseries{Pred}    & \bfseries{Pred}     & 3.89 $\pm$ 0.11   & 3.95 $\pm$ 0.05 \\
\bfseries{1/2 GT}  & \bfseries{GT} & 3.83 $\pm$ 0.04   & 3.89 $\pm$ 0.10 \\
\bfseries{Mix}     & \bfseries{Mix}    & 3.93 $\pm$ 0.06   & 3.98 $\pm$ 0.05 \\
\bottomrule  
\end{tabular}
\caption{
Results of SVS. GT refers to the ground truth, Pred indicates the results from our ASA model, and Mix represents a mix where half of the data is automatically annotated and the other half is ground truth.
}
\label{tab: svs}
\end{table}
To further validate STARS's effectiveness, we use STARS to annotate GTSinger's Chinese part and our data, and employ the latest style-controllable SVS model TCSinger\citep {zhang2024tcsinger} for generation tasks. We use MOS-Q for quality and naturalness assessment, and MOS-C for expressiveness evaluation of the generated singing's style control.

As shown in Table~\ref{tab: svs}, the following conclusions can be drawn: By comparing the first two rows, where training is performed on ground-truth, we observe that using our model's annotations yields MOS-C and MOS-Q scores nearly identical to those obtained with ground-truth annotations. Similarly, comparing the second and third rows, which correspond to training with ground-truth versus our model's annotations, and evaluating with our predicted results, shows only minimal differences in MOS-C and MOS-Q scores. These findings indicate that training exclusively with our annotated data achieves performance comparable to using ground-truth annotations, demonstrating the effectiveness of our fully automated annotation model.
Furthermore, comparing the last two rows—where training is conducted with half ground-truth data versus a mixture of half ground-truth and half predicted annotations—reveals an improvement in performance. This suggests that augmenting the dataset with our model's predictions can effectively enhance overall SVS model performance.

\section{Conclusion}
In this paper, we introduce STARS, the first unified framework for Singing Transcription, Alignment, and Refined Style Annotation. We construct a multi-level framework that efficiently extracts audio features at five granularities—frame, word, phone, note, and sentence—using a hierarchical acoustic encoder. Our approach enables a complete automatic singing annotation pipeline, sequentially performing singing Lyric alignment, note transcription and alignment, phone-level technique prediction, and global style prediction. Experimental results demonstrate that our model achieves high performance across all sub-tasks, and the annotated outputs are further validated in a singing synthesis task, confirming the effectiveness of our approach.

\section{Limitations}
Our model has two main limitations. First, it currently only classifies global style attributes of the singing voice and generates simple captions using predefined templates. In the future, we aim to enhance this capability by connecting the model to large language models, enabling more dynamic and context-aware caption generation.
Second, the model has been validated solely on Chinese and English datasets. Further validation on datasets from other languages is necessary to assess its generalizability. In future work, we plan to extend the model to singing data in additional languages.

\section{Ethics Statement}
The automatic singing annotation model may be subject to misuse in the following ways.
The model could be used to generate synthetic singing voices that closely resemble real individuals or artists without their consent, potentially leading to concerns around authenticity and intellectual property.
As the model is trained primarily on English and Chinese datasets, its performance on other languages or diverse cultural contexts may be limited, potentially resulting in biased or inaccurate annotations for non-target languages or dialects.
To mitigate these issues, we encourage transparent usage, proper attribution, and the continued development of ethical guidelines for synthetic media generation. Additionally, we also plan to explore the ways to make the model more inclusive and adaptable to a broader range of languages and contexts.
\section*{Acknowledgement}
This work was supported by National Natural Science Foundation of China under Grant No.62222211 and National Natural Science Foundation of China under Grant No.U24A20326.

\bibliography{custom}

\appendix
\section{Details of Model}
\subsection{Architecture}
\label{sec: app2model}
The model input for STARS consists of Mel spectrograms and the extracted f0, which are encoded separately using two encoders, denoted as $\text{E}_{\text{M}}$ for the Mel spectrogram and $\text{E}_{\text{P}}$ for f0. The Mel encoder is constructed from linear layers and residual convolution blocks, while the f0 encoder consists of an embedding layer. The fused features are then fed into five levels of an acoustic encoder to extract audio features at different hierarchical levels: frame, word, phone, note, and sentence.

Each acoustic encoder begins with residual convolution blocks followed by a CMU Encoder for feature extraction. The encoder and decoder of the U-Net backbone include four downsampling and upsampling layers, with a downsampling rate of $16 \times$. The U-Net module is further enhanced by two layers of Conformer blocks, where the FeedForward layers in each Conformer are replaced with FreqMOE layers. Each FreqMOE contains four experts, and the input features are equally split into four parts, each processed by an expert. The outputs of the four experts are then concatenated.

After feature extraction from the CMU Encoder, the sentence-level features are obtained by averaging across all frames and then they are expanded. For frame-level features, no further processing is performed. For the phone, word, and note levels, average pooling is applied based on boundary information. These pooled features then pass through convolutional layers followed by a Vector Quantization layer with a codebook size of 128 for feature filtering. Finally, the Length Regulator is applied to expand the features according to the corresponding boundary.

The Global Style Predictor is composed of two layers of cross-attention, which predict the type of each class. The overall architecture parameters are shown in Table~\ref{tab: arch}.

\begin{table}[t]
\centering
\setlength{\belowcaptionskip}{-0.4cm}
\begin{tabular*}{\hsize}{l|c|c}
\toprule
\multicolumn{2}{c|}{Hyperparameter}         & Model    \\
\midrule
\multirow{3}*{\shortstack{Mel\\Encoder}}    & Conv Kernel   & 3 \\
~                                           & Conv Layers   & 2 \\
~                                           & Hidden Size   & 256   \\
\midrule[0.2pt]
\multirow{3}*{\shortstack{Condition\\Encoder}}          & Pitch Embedding           & 300     \\
~                                                       & UV Embedding              & 3     \\
~                                                       & Hidden Size            & 256   \\
\midrule[0.2pt]
\multirow{2}*{\shortstack{Vector\\Quantization}}    & Code Num        & 128     \\
~                                                   & Hidden Size            & 256   \\
\midrule[0.2pt]
\multirow{4}*{\shortstack{U-Net}}                       & Kernel Size                  & 3     \\
~                                                       & Enc \& Dec Layers         & 4     \\
~                                                       & Downsampling Rate         & 16   \\
~                                                       & Hidden Size         & 256   \\
\midrule[0.2pt]
\multirow{6}*{\shortstack{FreqMOE\\Conformer}}                   & Kernel Size               & 9     \\
~                                                       & Head Num                    & 4     \\
~                                                       & Layers                    & 2   \\
~                                                       & Attention Hidden          & 256   \\
~                                                       & MOE Hidden                & 256   \\
~                                                       & Expert Num                & 4   \\
\bottomrule
\end{tabular*}
\caption{
Hyperparameters of STARS.
}
\label{tab: arch}
\end{table}

\label{sec: appendix4data}
\begin{table*}[t]
\centering
\small
\begin{tabular}{l|c|ccccccccc|c}
\toprule
\multirow{2}{*}{\bfseries{Setting}} & \multirow{2}{*}{\bfseries{Metric}} & \multicolumn{10}{c}{\textbf{Technique Prediction}} \\
& & {BUB} & {BRE} & {PHA} & {VIB} & {GLI} & {MIX} & {FAL} & {WEA} & {STR} & \bfseries{TEC} \\
\midrule
\multirow{2}{*}{Bilingual} 
& F1   & 74.0 & 50.5 & 78.6 & 50.0 & 69.7 & 87.5 & 86.6 & 94.9 & 99.6 & 76.8 \\
& ACC  & 98.5 & 89.0 & 94.6 & 96.3 & 84.2 & 81.1 & 91.1 & 90.7 & 97.3 & 91.4 \\
\bottomrule
\end{tabular}
\caption{
The objective results of phone-level technique prediction on the bilingual dataset. }
\label{tab: bitech}
\end{table*}

\begin{table}[t]
\centering
\small
\vspace{2mm}
\begin{tabular}{l|ccccc}
\toprule
\bfseries{setting} & \textbf{EMO} & \textbf{PAC} & \textbf{RNG} & \textbf{LAN} & \textbf{GEN} \\
\midrule
\bfseries{Single}  & 52.5 & 71.3 & 59.0 & - & 100.0 \\
\bfseries{Bilingual}  & 48.6 & 71.8 & 76.9 & 100.0 & 100.0 \\
\bottomrule
\end{tabular}
\caption{
The results of the global style detection. EMO refers to emotion, PAC to pace, RNG to pitch range, LAN to language, and GEN to gender.
}
\label{tab: style}
\end{table}

\subsection{FreqMOE}
\label{sec: moe}
Following previous audio generation systems with MOE \citep{zhang2025versatile,zhang2025isdrama}, the FreqMOE (Frequency Mixture of Experts) module is designed to enhance the representation of input features by leveraging multiple experts, each processing a distinct subset of the input. The module operates by splitting the input feature map into multiple chunks and passing each chunk through a separate expert network. The outputs from all experts are then concatenated to form the final representation. Specifically, the FreqMOE can be expressed as:

\begin{equation} 
\text{FreqMOE}(\mathbf{X}) = \text{Concat}_{k=1}^K E_k(\mathbf{X}^{(k)}), 
\end{equation}

where $\mathbf{X}_k \in \mathbb{R}^{T \times D / K}$ represents the $k$-th chunk of the input feature map, $\mathbf{X}$, split along its feature dimension. $E_k$ denotes the $k$-th expert network, which processes $\mathbf{X}_k$. 

\section{Audio-Phoneme Alignment}
\label{sec: appalign}

In this appendix, we briefly describe our alignment algorithm that synchronizes a sequence of phoneme labels with the corresponding audio (or video) frames. The algorithm is based on dynamic programming and is inspired by Viterbi decoding, which efficiently finds the most likely alignment path through a state-space representing both phoneme and silence (or blank) predictions.

\subsection{Overview}
Given an audio signal, a neural network produces frame-level log-probabilities for phoneme classes as well as for silence. In addition, a boundary detection mechanism provides probabilities indicating the likelihood of transitions between phoneme segments. The alignment problem is then formulated as finding the optimal path that maximizes the overall likelihood, taking into account both the phoneme content and the temporal boundaries between phonemes.

\subsection{Dynamic Programming Formulation}
Let $T$ be the number of time frames and $L$ be the number of phonemes in the target sequence. We define a score matrix $\mathbf{D}\in\mathbb{R}^{T \times (2L+1)}$, where each column corresponds to a state representing either a phoneme or an interleaved blank state. The algorithm initializes $\mathbf{D}$ with scores computed from the first frame's predictions and then iteratively updates the matrix as follows:
\[
D(t,k) = \max \left\{
\begin{array}{ll}
D(t-1,k) + s_{k}^{(t)},\\[1ex]
D(t-1,k-1) + s_{k}^{(t)},\\[1ex]
D(t-1,k-2) + s_{k}^{(t)},
\end{array}
\right.
\]
Here, $s_{k}^{(t)}$ denotes the log-probability score at time $t$ for state $k$, which is computed based on the phoneme, silence, and boundary predictions. A corresponding backtracking matrix $\mathbf{B}$ is maintained to record the optimal transition at each step.

\subsection{Backtracking and Temporal Mapping}
After processing all frames, the optimal alignment path is recovered by backtracking through $\mathbf{B}$ starting from the final frame. This path indicates, for each time frame, the most likely state (i.e., phoneme or blank) that generated the observation. Finally, by mapping the indices of phoneme states to time instants—using the known hop size of the audio frames—the algorithm produces onset and offset times for each phoneme. This procedure allows us to effectively synchronize phonetic labels with the audio signal.

\subsection{Summary of the Algorithm}
The overall alignment procedure can be summarized as follows:
\begin{enumerate}
    \item \textbf{Initialization:} Set up the dynamic programming matrices $\mathbf{D}$ (for scores) and $\mathbf{B}$ (for backtracking) using the initial predictions.
    \item \textbf{DP Matrix Update:} For each time frame $t=1,\dots,T-1$, update the scores for all states by considering self-transitions, transitions from the previous state, or skips (modeling boundaries).
    \item \textbf{Backtracking:} Recover the optimal alignment path by backtracking through $\mathbf{B}$.
    \item \textbf{Temporal Mapping:} Convert the alignment indices to temporal onset and offset times for each phoneme using the audio frame hop size.
\end{enumerate}

\section{Details of Experiments}
\label{sec: appendix4}
\subsection{Dataset}
\label{sec: appendix4data}
The open-source singing dataset used in our experiments includes the Chinese and English subsets of GTSinger~\citep{zhang2024gtsinger}. We use the dataset under the CC BY-NC-SA 4.0 license.
For the recordings, we select one male and one female professional singer, each paid \$350 per hour, and they agree to make their contributions available for research purposes. During the recording sessions, the singers are instructed to apply and label the technique annotations at both the sentence and phoneme levels. Phoneme segmentation is refined using the Montreal Forced Aligner (MFA), with additional manual adjustments to ensure accuracy. The annotators are compensated at a rate of \$15 per hour for their work.

\subsection{Evaluation Metrics}
\label{sec: appendix4metr}
For lyric alignment, we use two objective metrics: Boundary Error Rate (BER) and Intersection Over Union (IOU) score. The Boundary Error Rate (BER) measures the proportion of misplaced boundaries within 20 ms tolerance distance. The IOU score is defined as the ratio of the duration of the overlapping segment between two notes to the duration of the combined time span covered by both notes.

For note transcription, we use the objective metric COnPOff (Correct Onset, Pitch, and Offset). COnPOff assesses whether both the boundaries and pitch of the note are correctly predicted. Additionally, we compute the Raw Pitch Accuracy (RPA) to evaluate the overall pitch prediction performance. RPA is calculated by transforming both the ground truth (GT) and predicted note events into frame-level sequences and computing the matching scores.

In our evaluation process, we employ STARS to exclude silent notes and designate the boundaries enclosing each note as its onset and offset. For onset and offset evaluation, the tolerance is set to 50 ms, with the offset tolerance being the larger of 50 ms or 20\% of the note duration. The pitch tolerance is set to 50 cents.

For the singing voice synthesis experiment, we randomly select 30 generated singing clips for subjective evaluation. Each generated sample is rated and evaluated by at least five professional listeners. In the MOS-Q evaluation, listeners focus on rating the quality and naturalness of the generated singing voice. In the MOS-C evaluation, they assess whether the generated singing matches the style of the given prompt. All scores are rated on a five-point scale. Each participant is paid \$10 per hour.

\subsection{Experiment}
\label{sec: appendix2biling}

As shown in Table~\ref{tab: bitech}, we observe that when evaluating on a bilingual dataset, the overall prediction accuracy for the nine singing techniques is relatively high, comparable to the results obtained from a single-language dataset. This demonstrates the feasibility of our model for multilingual datasets.

According to the results presented in Table~\ref{tab: style}, we find that attributes such as language, gender, and vocal range, which are relatively fixed across the entire singing performance, yield better prediction results. n contrast, the model performs less effectively on attributes like emotion and vocal range, which may vary across different segments of the song. Further analysis of the dataset reveals that annotations for these attributes are typically provided at the sentence level, whereas emotion and vocal range fluctuate within different sections of a song. This variability leads to a decrease in prediction accuracy for individual segments of the singing voice.

\end{document}